\documentclass[a4paper,11pt]{article}
\pdfoutput=1 

\usepackage{jheppub} 

\usepackage[T1]{fontenc} 
\usepackage{graphicx}
\usepackage{pifont}
\usepackage{cancel}
\usepackage{pifont}
\usepackage{mathalfa,amssymb}
\usepackage{bbm}
\usepackage{bbold}
\usepackage{amsmath}
\usepackage{amsthm}
\usepackage{xcolor}
\usepackage[english]{babel}
\usepackage{multirow}
\usepackage[toc,page]{appendix}
\usepackage{hyperref}

\title{\boldmath Magnetic charges in Supergravity }

\author[a]{Bilyana L. Tomova}

\affiliation[a]{DAMTP, Centre for Mathematical Sciences, University of Cambridge,\\ Wilberforce Road, Cambridge,  CB3 0WA, U.K. }

\emailAdd{bt363@cam.ac.uk}

\abstract{In this paper we study the dual charges of $\mathcal{N}=1$ supergravity in asymptotically flat space-time. The action considered is the usual supergravity action with a topological contribution. This is the Nieh-Yan term and the magnetic term of the free Rarita-Schwinger field. Through methods of the covariant phase space formalism we construct the charges conjugate to supersymmetry, diffeomorphism and Lorentz transformations. The additional term in the action will lead to new, non-vanishing contributions to these charges. The magnetic diffeomorphism charges are equivalent to the ones previously found for gravity, while the dual supersymmetric charges are new and do not appear for the free Rarita-Schwinger field. The dual Lorentz charges serve to regularize the previous two. We find that the asymptotic symmetry group for supergravity can only include globally well-defined super-rotations.
}

\begin{document} 
\maketitle
\flushbottom

\section{Introduction}
\label{sec:intro}

The usual Maxwell equations that govern the behavior of the electric and magnetic fields, seemingly forbid the existence of magnetic monopoles. This conclusion was first challenged by Dirac. His insight was that by choosing a special couple of gauge potentials, that are not everywhere continuous, and removing the origin of the coordinates, one can construct a magnetic field with a monopole.

Later, in the study of Yang-Mills theory it was shown that one can add an additional gauge and Lorentz invariant term to the action, with a coupling constant $\theta$ (hence the name theta term). In special cases where the coupling constant varies with the background, one can have magnetic field, seemingly caused by monopole. This happens, for example, when one has electromagnetic radiation propagating inside of a topological insulator ($\theta  = \pi $) and outside ($\theta = 0 $).

There are a lot of parallels between Yang-Mills theory and the Einstein theory of gravity. The study of dual charges is no exception. In the fifties and sixties, a very peculiar solution to the vacuum Einstein equations was discovered - the Taub-NUT metric \cite{12}, \cite{13}. This solution is characterized by the NUT parameter, which can be viewed as a gravitomagnetic monopole \cite{25} - the dual charge of the gravitational mass. In recent years it has been shown \cite{27}, that the presence of the NUT charge requires that the subleading (with respect to the radial coordinate) components of the angular metric (written in the Bondi gauge \cite{1}), is not continuous on the sphere. What is more,  this part of the metric transforms as a gauge field under the action of the asymptotic symmetry group \cite{14}. If one chooses this field to be the gauge potential of the Dirac monopole, one obtains the the Taub-NUT metric \cite{14}.

On the other hand, the study of asymptotic symmetries of asymptotically flat space-times lead to the discovery of an infinite dimensional group of asymptotic charges \cite{20}. They represent different modes of the mass and angular momentum flux. This begs the question whether an infinite number of gravitational magnetic charges exist. Indeed these charges were found \cite{14}, \cite{15}. Subsequently, their existence was better understood in the Hamiltonian formalism of gravity \cite{16}.  Adding an additional term to the Einstein-Palatini action, one can derive these charges via the covariant phase space formalism. This additional term is the Holst term. It is similar to the theta term in Yang-Mills theory and does not change the equations of motion. Indeed one can derive the magnetic Yang-Mills charges in a similar fashion \cite{39}. If one couples gravity to fermions, the Holst term needs to be replaced by the Nieh-Yan term \cite{16}. 

In this paper we extend these efforts to supergravity. The topological term in the action will have two contributions - the Nieh-Yan term and the magnetic term from the free Rarita-Schwinger field \cite{24}. We choose the metric to be asymptotically flat and we write it in Bondi coordinates. The usual asymptotic symmetry group on this metric is the BMS. Here it would be enlarged to in two ways. Firstly, because of supersymmerty it will be extended to the super-BMS group \cite{26}. Furthermore, it will have magnetic charges as well as the usual super-BMS charges.

\section{Symplectic structure} \label{2}

In order to study the normal and dual Hamiltonian charges for supergravity we first need to set the tools of this analysis. Therefore, in this section, we briefly show how to construct the phase space for a gauge theory. The phase space for a field is an infinite-dimensional symplectic manifold, where every point is a specific field configuration. These configurations are constrained by the equations of motion and the boundary conditions. The conserved charges of the theory are functions on phase space. Mathematically they are are expressed as functionals of the fields. Their commutator relationship and infinitesimal variations are defined via the symplectic form. The steps for defining this symplectic form are straightforward  \cite{3}, \cite{17}. 

\newpage 
First we vary the Lagrangian,

\begin{gather}\label{3.1}
\delta L (\phi, \delta \phi ) \approx d \Theta (\phi, \delta \phi )
\end{gather}
Here and throughout this paper the sign $\approx$ means equality on shell.
The boundary term $\Theta $ is the pre-symplectic potential. The pre-symplectic density and form are defined respectively as,

\begin{flalign}\label{3.2}
& w (\phi, \delta _1 \phi, \delta _2 \phi ) = \delta _1 \Theta (\phi, \delta _2 \phi ) -  \delta _2 \Theta (\phi, \delta _1 \phi )\\
& \tilde{ \Omega }  (\phi, \delta _1 \phi, \delta _2 \phi ) = \int _{\Sigma }  w (\phi, \delta _1 \phi, \delta _2 \phi )
\end{flalign}
where $\Sigma $ is a Cauchy surface in the space-time manifold. The pre-symplectic form will be in general degenerate on the phase space. Its degenerate directions correspond to transformations, that are not physical, but are rather redundant descriptions of the same field configuration. If we quotient the phase space by the degenerate directions, we get the pre-symplectic form $\Omega $. The infinitesimal variation of a Hamiltonian charge, conjugate to a field variation $\delta _{\chi } \phi $ is

\begin{flalign} \label{dH}
&  \cancel{\delta} H   _{\chi } [\phi ] =  \int _{\Sigma }  w (\phi, \delta _{ \chi} \phi, \delta  \phi )
\end{flalign} 

For gauge transformations this can be re-written as a boundary term \cite{17}. This variation is not in general an exact one form on phase space. This can be due to two to reasons- the transformation is not canonical, or there is a charge flux through the boundary of the space-time region. In the second case, one can impose the condition that the variation of the field on the boundary is zero and thus make the Hamiltonian charge integrable \cite{3}.

The commutator of two integrable charges is defined as \cite{17},

\begin{flalign}
&[ H_{\chi }, H_{\xi } ] [\phi ]= \delta _{\xi } H _{\chi } [\phi ] =  \int _{\Sigma } w (\phi, \delta _{\chi } \phi , \delta _{\xi } \phi )  = H_{[\xi, \chi ]} +K  _{\xi , \chi } [\bar{\phi}] \\
& K _{\xi, \chi } = \int _{\Sigma } w (\bar{\phi} , \delta _{\chi } \bar{\phi} , \delta _{\xi } \bar{\phi} )  + N _{[\xi  , \chi ]} [\bar{\phi}]
\end{flalign}

Where $K $ is the central extension and $\bar{\phi} $ is a reference field configuration, which is the starting point of integration of the charges on phase space.The term $N $ is a reference charge.

\section{The action} \label{section1}

The action of gravity with magnetic term, coupled to fermions has been derived in \cite{16}. It is the usual Einstein-Palatini action plus the Nieh-Yan term. The magnetic charges of the free massless Rarita-Schwinger field were studied in \cite{24}. Here, we will combine these efforts and study the magnetic charges of the $\mathcal{N} =1$ supergravity theory.

We work in the 1.5 order formalism \cite{22}. Initially, we consider the that the frame field, spin connection and gravitino field are all independent variables. By varying the action with respect to the spin connection, we can determine its expression in terms of the two other fields. The presence of spinor field, gives rise to a non-vanishing torsion. Once the expression for the spin connection is determined, we put it back in the action and proceed as if the only independent variables are the frame and gravitino field. For more details of this formalism see \cite{22}.

The usual action for $\mathcal{N}=1 $ supergravity in the Einstein-Palatini formalism is the following,

\begin{flalign}
&S_{SG} =  \frac{1}{2\kappa ^2} \int _{\mathcal{M}}  \,   \varepsilon _{abcd} \, e^{a } \wedge e^{b } \wedge R^{ c d  } - \bar{\psi} \wedge \gamma _5 \gamma \wedge  \nabla \psi
\end{flalign}

Here and throughout this paper, Greek letters $\mu, \nu$ are space-time indices, and Latin letters $a ,b$  are tangent space indices.

The topological term, that will give raise to magnetic charges is, 

\begin{flalign}
S_{\ast SG} = - \frac{i \lambda }{2\kappa ^2} \int _{\mathcal{M} }  \  e_a \wedge e_b \wedge R ^{ab} - T^a   \wedge T_a  - d \left( \bar{\psi } \wedge \gamma \wedge \psi \right)
\end{flalign}
where $T^a$ is the torsion, defined as $de^a  + e^b \wedge \omega _b ^{\ a } = T^a $. The first part is the topological term for gravity coupled to fermions  \cite{16}. It is equall to $ e_a \wedge e_b \wedge R ^{ab} - T^a \wedge T_a = d( e_a \wedge T^a)$. Re-writing it like this however, hides away important terms, that are technically zero, but whose contribution to the pre-symplectic potential is not.
The action is invariant under supersymmetry, diffeomorphism and local Lorentz transformations \cite{22}. These transformations act respectively on the frame and gravitino field in the following way,

\begin{flalign}
& \delta _{\epsilon } e^a _{\mu }  =\frac{1}{2} \bar{\epsilon } \gamma ^a \psi _{\mu} \quad \delta _{\epsilon } \psi _{\mu } =  \nabla _{\mu } \epsilon \\
& \delta _{\xi } e^a  _{\mu }= \mathcal{L }_{\xi } e^a _{\mu } = \xi ^{\nu } \partial _{\nu } e^a _{\mu } + e^a _{\nu } \partial _{\mu } \xi ^{\nu } \quad  \delta _{\xi } \psi  _{\mu }= \mathcal{L }_{\xi } \psi _{\mu } = \xi ^{\nu } \nabla _{\nu } \psi  _{\mu } + \psi _{\nu } \nabla _{\mu } \xi ^{\nu }  - \frac{1}{4} \nabla _{\rho } \xi _{\nu }  \gamma ^{\rho } \gamma ^{\nu } \psi _{\mu} \\
& \delta _{\Lambda } e^a = \Lambda ^a _{\ b } e^b \quad \delta _{\Lambda } \psi  = -\frac{1}{4} \Lambda _{ab} \gamma ^{ab} \psi 
\end{flalign}
where $\epsilon$ is the gauge spinor, $\xi $ is the vector field, generating diffeomorphism, and  $\Lambda $ generates a local Lorentz transformation.

The variation of the action is,

\begin{flalign}
\delta S  = &\frac{1}{2\kappa ^2} \int _{\mathcal{M}}   \left( 2 \varepsilon_{abcd} e^b \wedge R^{cd}  - \bar{\psi} \wedge \gamma _5 \gamma_a  \nabla \psi   \right) \wedge \delta e^a \\
&+\left( 2  \varepsilon_{abfd}\, de^a \wedge e^b  +2  \varepsilon_{abcd} e^a \wedge e^b \wedge  \omega _{\ f} ^{ c } - \frac{1}{4} \bar{\psi} \wedge \gamma _5 \gamma \gamma _{fd} \wedge  \psi\right) \wedge \delta \omega ^{fd}\\
&- \left( \nabla \bar{\psi}  \gamma _5 \wedge \gamma  \right) \wedge \delta \psi \\
& + d \left(   e_{a b c d} e^a \wedge  e^b \wedge \delta \omega^{ cd } - \bar{\psi } \gamma _5 \wedge \gamma \wedge \delta \psi  \right)\\
& - i \lambda  \left(  e_a \wedge e_b \wedge \delta \omega ^{ab}  -2 \delta e^a \wedge T_a  - \delta \left( \bar{\psi } \wedge \gamma \wedge \psi \right) \right)
\end{flalign}

In the first line we expressed the variation of the action with respect to the frame field - this gives the Einstein equation. The second line is variation with respect to the spin connection  and it gives the equation for the torsion and contorsion tensor,

\begin{flalign} \label{connection}
& \omega _{\mu ab} =  \omega _{\mu ab} (e) + K_{\mu ab } (\psi ) \quad  K_{\mu [ a b  ] } (\psi ) = \frac{1}{4} \left( \bar{\psi } _b \gamma _a   \psi _{\mu}  -  \bar{\psi } _a \gamma _b   \psi _{\mu}   - \bar{\psi } _a \gamma _{\mu}    \psi _b \right) \\
& T_ { [ \mu \nu  ] b } (\psi  ) = \frac{1}{2} \bar{\psi } _{\mu} \gamma _b   \psi _{\nu}  
\end{flalign}

The third line is the equation of motion for the gravitino field. The two final lines are pre-symplectic potential for $\mathcal{N} =1$ supergravity with magnetic term,

\begin{flalign}
\Theta  = & \frac{1}{2 \kappa ^2} \left(\varepsilon_{a b c d} e^a \wedge  e^b \wedge \delta \omega^{ cd } -  \bar{\psi } \wedge  \gamma _5  e^a \gamma _a \wedge  \delta \psi \right)  \\
&-\frac{ i \lambda }{\kappa ^2 } \underbrace{\left( e_a \wedge e_b \wedge \delta \omega ^{ab}  -2 \delta e^a \wedge T_a - \delta( \bar{\psi } \wedge \gamma \wedge \psi ) \right)}_{\Theta ^{\ast }}
\end{flalign}

Before we proceed with studying the charges we note something useful about the magnetic piece in the symplectic density. 

\begin{flalign}
&\Theta ^{\ast } (\delta ) = \delta ( e_a \wedge e_b \wedge \omega ^{ab}) - 2 \delta e^a \wedge de _a - \delta( \bar{\psi } \wedge \gamma \wedge \psi ) \\
& \Omega ^{\ast }( \delta _1 , \delta _2) = 2d(\delta _1 e^a \wedge \delta _2 e_a) = d\delta _1( e^a \wedge \delta _2 e_a)  - d\delta _2( e^a \wedge \delta _1 e_a) 
\end{flalign}

Because $\delta $ is the exterior derivative on phase space.
\subsection{Asymptotic conditions}

The equations of motion, that the fields $e^a$ and $\psi $ need to satisfy are already given in section \ref{section1}. In this section, we introduce the boundary conditions of the fields. Firstly, we demand that the metric is asymptotically flat. In the Bondi gauge \cite{1}, \cite{6}, \cite{33} an asymptotically flat metric takes the following form,

\begin{flalign}\label{metric}
 &ds ^2 =  - e ^{2\beta}f^2 du ^2 - 2 e^{2\beta} dudr +r^2 h_{AB} (dx ^{A} - U^A du )(dx ^{B} - U^B du) \\
& h _{A B}=  \gamma _{AB} +\frac{C _{AB}}{r}  + \frac{d _{AB}}{r^2}  + O(r^{-3}) \quad f ^2 (u, r, x^A)= 1 - \frac{2 M (u, x^A )}{r} +O(r^{-1})\\
 & e ^{2\beta } =1+O (r^{-2} ) \quad g_{uA} = \frac{1}{2} D_B C^{AB} + \frac{1}{r} \left(  \frac{1}{4} C_{BC} D_A C^{CB} + N_A\right) \quad d_{AB} = \frac{1}{4}\gamma _{AB} C_{CD} C^{CD}
\end{flalign}

Here $D_A$ denotes the covariant derivative with resepect to the metric of the unit two-sphere $\gamma _{AB}$,  $M(u, x^A)$ is the Bondi mass, that can vary with time and $N_A$ is the angular momentum. The subleading tensor $C_{AB}$ characterizes a gravitational wave and is related to the time derivative of the mass. Residual gauge freedom can be used to make it traceless and to set the determinant  $\det (g_{AB}) $ to be $r ^2 \det (\gamma _{AB})$. 

Since we work in the Einstein-Palatini formalism, the metric is not the gravitational field of interest. Rather we need the frame fields $e^a _{\mu }$, which satisfy $e^a _{\mu } e^b _{\nu } \eta _{ab } = g_{\mu \nu }$, where $\eta _{ab}$ is the Minkowski metric. The frame fields are defined up to a local Lorentz transformation. Our choice for them is,

\begin{flalign}
&e^0 = \frac{e^{\beta }}{f}dr + e^{\beta } f du\quad 
e^1=  \frac{e^{\beta }}{f}dr  \\
&e ^i =  r E ^i _A \left ( dx^A   - U^A du \right) \quad E ^i _A  E^j _B \delta _{ij} =h _{AB} \quad i, j \in \{2, 3 \}
\end{flalign}
Any variation should preserve the leading terms of these fields.

The gauge condition for the gravitino field is,

\begin{flalign}\label{fgauge}
\gamma ^{\mu } \psi _{\mu } = 0
\end{flalign}

This simplifies the equations of motion for $\psi$. The boundary conditions on $\psi $, will be dictated by the fact that supersymmetric transformations must preserve the gauge and asymptotic behavior of the metric,

\begin{flalign}
\delta _{\epsilon } g_{rr} = \bar{\epsilon } \gamma _r \psi _r = 0 \quad \delta _{\epsilon } g_{ur} =\frac{1}{2} \bar{\epsilon } \gamma _{(u} \psi _{r)} = O(r^{-2}) \quad \delta _{\epsilon } g_{rA} =\frac{1}{2} \bar{\epsilon } \gamma _{(r} \psi _{A)}  = 0
\end{flalign}

To leading order this and the fermionic gauge condition are satisfied when,

\begin{flalign}\label{conpsi}
&\epsilon \sim O(1) \quad \psi _A \sim O(1) \quad \psi _u \sim  O(r^{-1}) \quad \psi _r \sim O(r^{-2}) \\
&\gamma ^u \psi_r = 0 \quad  \gamma ^{u(0)} \psi _A ^{0} = 0   \quad \gamma ^{u(0)} \psi _u ^{(-1)} = 0  \quad \gamma ^{A(-1)} \psi _A=0
\end{flalign}

Notice that, because the matrix $\gamma ^u $ is nilpotent the equations of the second line do not imply that $\psi _{r }$, $\psi _A ^{(0)}$ or $\psi _u ^{(-1)}$ are 0. The justification for these boundary conditions are the following. Firstly, $\epsilon \sim O(1)$ comes from the fact that the commutator of the supersymmetric transformations of the metric, should be a diffeomorphism. This is better explained in the next section. Next, because $\gamma _{AB}$ is fixed $\psi _A$ should be $O(1)$. It cannot be more sub leading, because the subleading tensors of $h_{AB}$ depend on $C_{AB}$. The trace-free condition requires that $\gamma ^{A(-1)} \psi _A ^{(0)}  = 0$. Combining this with $ \delta _{\epsilon } g_{rA} $, we can reach the conclusion that $\psi _r ^{(-1)} = 0$. Last but not least, remembering that the supersymmetric variation, should also respect the linearized equations of motion for the metric, we must also have to the following constraint on $\psi _{A} ^{(-1)}$,

\begin{flalign}\label{psiA-1}
\gamma _{(A} ^{(0)} \psi _{B)} ^{(0)} + \gamma _{(A} ^{(1)} \psi _{B)} ^{(-1)}  = \frac{1}{2}\gamma _{AB}C^{DC} \gamma _{(D} ^{(1)} \psi _{C)} ^{0} \ \Rightarrow \ \gamma ^{A(-1)} \psi _A ^{(-1)} = 0
\end{flalign}

Finally it is important to discuss what variations are allowed on the gravitino field. Super-translations preserve the leading order components of the field, while super-rotations do not. From super-symmetric transformation we require that $\delta _{\epsilon } \psi _A = O(r^{-1})$ and $\delta _{\epsilon } \psi _u = O(r^{-2})$. This is discussed in detail in section \ref{4.3}. Finally, for integrability of the supersymmetric charges we require that any other non-specific variation $\delta \psi _A$ is subleading. The fact that super-rotations violate this condition will result in a contribution to the central charge.
\section{Charges}

\subsection{Diffeomorphism charges}

The diffeomorphisms that respect the boundary and gauge conditions of the fields form the BMS group. The BMS group is generated by vector fields, that take the following form \cite{1}, \cite{6},

\begin{flalign}
&\xi ^u  = f(x^A) + \frac{u}{2} D_B f^B (x^A) \quad \xi ^A = f^A  - \partial _B \xi ^u \int \frac{1}{r^2}e^{2 \beta } h^{AB} dr \\
& \xi ^r = - r \partial _u \xi ^ u  + \frac{1}{2} \nabla^r \xi ^u - \partial _B \xi ^u \int U^B dr \quad \partial _u \xi ^u  = D_A f^A  
\end{flalign}

These generators can be divided in two subcategories - super-translations and super-rotations. The super-translation generators depend only on an arbitrary function on the sphere $f(x^A)$. When this function is taken to be the lowest order spherical harmonics, the transformation reduces to ordinary translation. Similarly, the super-rotation generators depend only on a vector field on the two sphere $f^B(x^A)$ and includes the Lorentz subgroup.  

In order to calculate the diffeomorphism charge, we use the fact that the variation of any quantity is just its Lie derivative, and we can use Cartan's magic formula,

\begin{flalign}
\Omega(\delta , \delta _{\xi} ) \approx \, &  \delta \Theta ( \delta _{\xi }) - d  \iota _{\xi} \Theta (\delta  ) - \iota _{\xi } \delta L = \delta \left(  \Theta ( \delta _{\xi })  - \iota _{\xi } L   \right)  - d  \iota _{\xi} \Theta (\delta) \\
& \cancel{\delta } H _{\xi }  =  \int _{\Sigma }  \delta \left(  \Theta ( \delta _{\xi })  - \iota _{\xi } L   \right)  - \int_{\partial \Sigma }  \iota _{\xi} \Theta (\delta) 
\end{flalign}

We see that the Hamiltonian charge, naturally splits into integrable and non-integrable part. The latter quantifies the flux of the given charge through the Cauchy surface, which here is chosen to be future null-infinity. At the past boundary of $\mathcal{I} ^+ $ this flux is set to be zero, by choosing appropriate $u$ fall-off conditions for gravitational radiation.

The calculations for the Hamiltonian charge density  are in appendix \ref{D}. The formula for the charge is,

\begin{flalign}\label{diffcharge}
\cancel{\delta } H  _{\xi }    +  i \lambda \cancel{\delta } \tilde{H}= \frac{1}{2 \kappa ^2 }  \int _{\partial \Sigma } &\delta\left(   \varepsilon _{a b c d}  e^a \wedge  e^b \, \iota_{\xi }\omega^{ cd }   -  \bar{\psi } \wedge \gamma _{5} \gamma \, \iota _{\xi }  \psi \right)   \\
&- \iota _{\xi} \left(  \varepsilon _{a b c d } e^a \wedge  e^b \wedge \delta \omega^{ cd }  -  \bar{\psi } \wedge \gamma _{5} \gamma \wedge  \delta  \psi \right)  + i \lambda (\mathcal{L}_{\xi } e^a \wedge \delta e_a)
\end{flalign}

First we note, that due to the boundary conditions on the fermionic field \ref{conpsi}, the part of the charge, that depends on $\psi $, is finite and it vanishes for super-translations. It is integrable for both super-translations and super-rotations. The integrability crucially relies on $\gamma ^{A(-1)} \psi _A ^{0} = \gamma ^{u(0)} \psi _{A} ^{(0)} = 0$. We see that in the context of supergravity, the finiteness of the global Lorentz generators of the gravitino field is a direct consequence of the asymptotic flatness of the metric. One does not need to make additional assumptions for the asymptotic behavior of the field, as is the case for the free Rarita-Schwinger field \cite{34}. 

The rest of the above expression is the usual diffeomorphism charge of pure gravity. The charge has normal and magnetic part. More specifically, comparing with the results from \cite{6}, \cite{16} and \cite{32}, the super-translation and super-rotation charges are respectively,

\begin{flalign}
H_{ST} + i \lambda \tilde{H} _{ST} =& \frac{1}{2 \kappa ^2} \int _{ \mathcal{I} ^+ _- } \sqrt{\gamma } \, f M   + i \lambda \int_{\mathcal{I} ^+ _-}  -  C_{B [ D} D _{A]} D^B \xi ^u \\
H_{SR}  + i \lambda \tilde{H} _{SR}=& \frac{1}{2 \kappa ^2} \int_{ \mathcal{I} ^+ _- } \sqrt{\gamma } \, -  \frac{1}{4} D_A f^A  C^2 + f^A N_A  + \bar{\psi }_{[A} \gamma _{5} \gamma _{B]}  \psi _C f^C \\
+&  i \lambda \int_{ \mathcal{I} ^+ _- } - C_{B [ D} D_{A]} D^B \xi ^u  + \frac{1}{4} f^B C_{C[A} D_{|B|} C^{C} _{D]}- \frac{1}{4} f_{[A }D_{D ] } C^2
\end{flalign}
where $C^2  = C_{AB} C^{AB}$. 
\subsection{Lorentz charges}

In addition to diffeomorphisms, we also need to consider local Lorentz transformations. They do not affect the metric and therefore one expects that they are unphysical.
The frame field transforms under diffeomorphisms and local Lorentz transformations as,

\begin{flalign}
\delta e^{a} _{\mu } = \xi ^{\nu } \partial _{\nu }e^{a} _{\mu }  + e^{a} _{\nu } \partial _{\mu } \xi ^{\nu } + \Lambda ^a _{\ b } e^{b} _{\mu }
\end{flalign}

If we take into consideration only the diffeomorphisms, the gauge and boundary conditions are not satisfied. We need local Lorentz transformations to regularize this. Their expressions up to leading order are,

\begin{flalign}
& \Lambda ^0 _{\ i }  =   - e^0 _{\mu } e^A _i  \partial _{A} \xi ^{\mu } \quad \Lambda ^1 _{\ i }  =   - e^1 _{r} e^A _i  \partial _{A} \xi ^{r } \\
&   \Lambda ^1 _{ \ 0 }   = -\frac{1}{f^2} \partial _r \xi ^r \quad  \Lambda ^i _{\ j} =  - e^A _j \mathcal{L} _{f^B} e^i _A + \delta _{\ j} ^i D_A f^A  
\end{flalign}

Notice that in the third line, $\delta _i ^j D_A f^A$  is the symmetric part of the term $- e^A _j \mathcal{L} _{f^B} e^i _A$, making $\Lambda _{ij}$ asymmetric as it should be. Furthemore, we observe that these transformations are needed to preserve  the boundary conditions, only if the BMS transformation in question is a super-rotation. Intuitively, this can be understood as the fact that super-rotations are a generalization of Lorentz transformations. Therefore, part of their effect on the frame field needs to be undone by a local Lorentz transformation.

We now proceed to study the charges related to these transformations. We know from \cite{16} that $\delta _{\Lambda } \left( \varepsilon _{abcd } e^a \wedge e^b \wedge \delta \omega ^{cd } \right) = 0 $. We also have,

\begin{flalign}\label{3.33}
\varepsilon _{abcd } e^a \wedge e^b \wedge \delta _{\Lambda } \omega ^{cd }=  d \mathcal{Q}_{\Lambda } + 2 \varepsilon _{abcd } T^a \wedge e^b \Lambda ^{cd}  = d \mathcal{Q}_{\Lambda } + 2 \psi \gamma _5 \gamma _{bcd } \wedge \psi \wedge e^b \Lambda ^{cd}
\end{flalign}

We know that the hamiltonian charge from $\mathcal{Q}$ in this setup is zero \cite{16}. From the fermionic part we have:

\begin{flalign}
& \delta _{\Lambda } \psi = -\frac{1}{4} \Lambda _{a b } (x) \gamma ^{ab} \psi  \quad \delta _{\Lambda } e ^b=  \Lambda (x)_{\ a} ^b e ^a \\
& \bar{\psi }  \wedge \gamma _5 \gamma \wedge \delta _{\Lambda } \psi  =  \bar{\psi }  \wedge \gamma _5 \gamma  \gamma _{cd} \wedge \psi \Lambda ^{cd}
\end{flalign}

Which simplifies the second term in \ref{3.33}. Furthermore, thanks to the identity \ref{5.8} (which works in the same way if one replaces $\omega _{ab}$ by $\Lambda _{ab}$), we can see that $\delta _{\Lambda } \left(  \bar{\psi }  \wedge \gamma _5 \gamma \wedge \delta \psi \right) = 0$. We conclude that the normal Lorentz charge vanishes.

On the other hand, the magnetic term from the Lorentz charges does not vanish. This seems bizarre at a first glance. However, this is actually useful as it serves to regularize the diffeomorphism magnetic charge.

\begin{flalign}
\left( \mathcal{L}_{\xi } e^a  + \Lambda ^a _{\ b } e^b \right) \wedge \delta e_a \sim O(1)
\end{flalign}

\subsection{SUSY charges}\label{4.3}
 
The charges conjugate to supersymmetry transformations are more complicated. We will study their expression in this section. First we work out the transformations of  the fields and the spin connection.

\begin{flalign}
\delta _{\epsilon }  e_{\mu } ^a =&  \frac{1}{2} \bar{\epsilon } \gamma ^a \psi _{\mu} \quad \delta _{\epsilon }  \psi _{\mu} = \nabla _{\mu } \epsilon   \\
 \delta_{\epsilon }   \omega _{\mu ab } (e) =&  \frac{1}{2} \partial _{\mu } \left( \bar{\epsilon } \gamma _{[a} \psi _{b]} \right) + e^{\rho } _a \partial _{\mu } \left(\bar{\epsilon } \gamma _b \psi _{\rho } \right)  + \frac{1}{2} \bar{\epsilon } \gamma _{[a } \psi ^{[\rho } e^{ \sigma  ] } _{b  ] } \partial _{\sigma } g_{\rho \mu} + \frac{1}{2} e^{\rho } _{[a} e^{\sigma } _{b] } \partial _{\rho} \left(  \bar{\epsilon} \gamma _{( \sigma} \psi _{\mu ) } \right) \\
\begin{split}\label{Ke}
 4 \delta _{\epsilon } K _{\mu ab } = &  \nabla _b \bar{\epsilon } \gamma _a \psi _{\mu } + \frac{1}{2} \bar{\epsilon} \gamma _b \psi ^{\nu } \bar{\psi }_{\nu } \gamma _a \psi _{\mu } + \bar{\psi }_b \gamma _a \nabla _{\mu } \epsilon \\
& - \frac{1}{2} \bar{\psi }_a \gamma ^c \psi _b \bar{\epsilon } \gamma _c \psi _{\mu } +  \nabla _a \bar{\epsilon } \gamma _{ \mu } \psi _{b} + \frac{1}{2} \bar{\epsilon} \gamma _a \psi ^{\nu } \bar{\psi }_{\nu } \gamma _{\mu } \psi _{b } - (a \leftrightarrow b )
\end{split}
\end{flalign}

Where $\epsilon $ is the gauge spinor. These transformations need to preserve the boundary and gauge conditions of the frame and gravitino field. This leads to various constraints. We first look at the gauge spinor,

\begin{flalign}\label{5.1}
[\delta _1, \delta _2 ] g_{\mu \nu }  = \nabla _{(\mu } \xi _{\nu)} \quad \xi_{\nu}  = \frac{1}{2} \bar{\epsilon} _1 \gamma _{\nu } \epsilon _2 
\end{flalign}

This looks like the transformation of the metric under diffeomorphism. We already know what conditions should be satisfied by the vector field $\xi ^{\mu }$ - it should generate a member of the BMS group. In particular  $\xi ^u $ should not be a function of $r$. This is achieved if the following property is satisfied by the gauge spinor $\epsilon$,

\begin{flalign}
\frac{e^{\beta }}{ f} \bar{\epsilon} _1 \left( \gamma _0 + \gamma _1 \right) \epsilon _2 \neq g(r) \ \Rightarrow \ \epsilon _i  = \sqrt{\frac{f}{e^{\beta }}} f_i (\theta , \phi )  + \gamma _r ^{(0)} \rho _i 
\end{flalign}

 From the above we see that $\epsilon ^{-1} = - \frac{M}{r} \epsilon ^ 0  + \frac{\gamma ^{u(0)}}{r} \rho ^{(-1)}$, (because $\gamma ^{u(0)} \times \gamma ^{u(0)} =0 $). From the gauge condition \ref{fgauge} at first and second order in $r$ the equations of motion for the gauge spinor are,

\begin{flalign}
 O(1) : & \gamma ^{u 0} \partial _u \epsilon ^{0}=0  \\\label{n1}
 O(r^{-1}) :  &  \gamma ^{A(-1)} \partial _A \epsilon ^{0}   + \frac{1}{2} \cot \theta \gamma ^2  \epsilon ^{0} + \gamma ^1 \epsilon ^0  +\gamma ^{u 0}\partial _{u} M \epsilon ^{0} +\gamma ^{u 0} \partial _u \epsilon ^{-1} =0 \\ \label{4.5}
&  \gamma ^{A(-1)} \partial _A \epsilon ^{0}   + \frac{1}{2} \cot \theta \gamma ^2  \epsilon ^{0} + \gamma ^1 \epsilon ^0  =0
\end{flalign}

This is the equation  for covariantly constant spinor and it agrees with the result from \cite{26}. It's expression is given in the appendix. This tells us that the function $\xi ^u = \frac{1}{2} \bar{\epsilon} _1 \gamma ^u \epsilon _2 $ is some linear combination of the lowest order spherical harmonics. Thus we recognize that the vector field $\xi ^{\mu }$  from \ref{5.1} generates only ordinary translations. The same conclusion was reached in \cite{35}. 
Notice that in order to obtain higher order  spin $1/2$ spherical harmonics, \cite{4}, and consequently more interesting super-translations, we would need a different constant multiplying $\gamma ^1$. This would mean setting $U^0 = \lambda \neq 1 $. However Einstein equations tells us that $U^0 =\frac{1}{2} R[\gamma]$. Knowing what the Ricci tensor is for the metric on the sphere, we can infer that this re-scaling would also re-scale the coefficients of $\gamma _{AB}$, so as to leave the above equation invariant. Therefore the equations of motion and the asymptotic flatness impose, that the commutator of two supersymmetric transformations can only be an ordinary translation. It would be interesting to see whether more interesting results can be obtained for other metrics. Finally, a quick calculation can show that by choosing $\rho ^{(-1))} =  - \gamma ^{0} M \epsilon ^0$, we can set $\delta _{\epsilon } \psi _u \sim O(r^{-2})$, which is going to be useful.

With these conditions in mind, we now proceed to the calculations of the fermionic charges, arising from supersymmetry. The calculations for the part of the charge, arising from the gravitino term in the action, are almost identical to the calculation for gauge charges of the free, massless Rarita-Schwinger field \cite{24},

\begin{flalign}
& \delta \Theta (\delta _{\epsilon })   -  \delta  _{\epsilon }\Theta (\delta ) \approx \delta \left( \Theta  (\delta _{\epsilon })-  I _{\epsilon } L \right)  - d I _{\epsilon } \Theta (\delta ) \\
&  \delta \left( \Theta (\delta _{\epsilon})  -  I _{\epsilon } L \right) \approx \delta \left( \Theta (\delta _{\epsilon}) \right) = \delta \left( \bar{\psi} \wedge \gamma _5 \gamma \wedge \nabla \epsilon \right) \\ \label{4.31}
& \approx d \left( \delta (\bar{\psi} \wedge \gamma _5 \gamma \epsilon)  +\bar{\psi} \wedge \gamma _5  T^a \gamma _a  \epsilon \right)  \\
& I_{\epsilon } \Theta (\delta ) = \bar{\epsilon } \gamma _5 \gamma \wedge \delta \psi
\end{flalign}
where the operator $I_{\epsilon }$ is defined as $I _{\epsilon } = \epsilon \cdot \frac{\delta}{\delta \psi}$. In the expression for $\delta  _{\epsilon }\Theta  $ I have ignored the term $\bar \psi  \wedge \gamma _5 \delta  _{\epsilon } \gamma \wedge \psi  $ because this is a four fermion term that vanishes thanks to the Fierz identity \ref{fierz}. Simiralry, $\bar{\psi} \wedge \gamma _5  T^a \gamma _a  \epsilon $ vanishes thanks to the cyclic identity for spinors \ref{cyclic}.

We now turn to the gravitational part of the symplectic form.

\begin{flalign}
\Omega (\delta, \delta _{\epsilon }) = \varepsilon_{a b c d} \int_{\Sigma}   \delta( e^a \wedge  e^b  \wedge \delta _{\epsilon }\omega^{ cd })  - \delta _{\epsilon } ( e^a \wedge  e^b \wedge \delta \omega^{ cd } )
\end{flalign}

First we note two things - $  \int_{\Sigma} \varepsilon_{a b c d}  e^a \wedge  e^b  \wedge B ^{ cd } = \int_{\Sigma} B_{\mu } ^{ \ \mu \nu } d \Sigma _{\nu} $, for any one-form with two antisymmetrized tangent space indices. Also, the connection form $\omega $ has two parts $\omega  = \omega (e) +K (\psi)$. From the expression of the supersymmetric variation of the contorsion tensor $K_{\mu a b }$ \ref{Ke} and the gauge condition \ref{fgauge}, we can conclude that  $ \varepsilon_{a b c d}  e^a \wedge  e^b  \wedge \delta _{\epsilon }  K^{ cd }=0 $. Any variation $\delta $ should preserve this gauge condition. Furthermore, most of the terms of  $\delta _{\epsilon} \left(\varepsilon_{a b c d}  e^a \wedge  e^b  \wedge \delta K^{ cd } \right) $ again vanish by the gauge condition. The remaining ones are of the form $ n_{\nu } \delta _{\epsilon} \bar{\psi} ^{\nu }  \gamma ^{\mu } \delta \psi _{\mu}$.  This will be sub-leading because we have set $\nabla _u \epsilon \sim O(r^{-2})$. We are left with the usual symplectic form of four-dimensional pure gravity. Its expression is,

\begin{flalign}
\Omega (\delta _1 ,  \delta _2) =& \frac{1}{2\kappa ^2} \int _{\mathcal{I} ^ + } \sqrt{\gamma} \, du d \theta d \phi \, \varepsilon^{AB} \, \varepsilon_{CD}  \, \delta_1\left( C_{A} ^D \right) \, \delta_2 \left( \partial _ u C^C _B \right) - (1 \leftrightarrow 2) \\
 \Omega (\delta , \delta_{\epsilon}) = & \frac{1}{2 \kappa ^2} \int _{\mathcal{I} ^ + }  \sqrt{\gamma} du d \theta d \phi \, \varepsilon^{AB} \, \varepsilon_{CD}  \, \bar{\epsilon} \gamma ^D  \psi _A \, \delta \left( \partial _ u C^C _B \right) - (\delta  \leftrightarrow \delta _{\epsilon })\\
=& - \frac{1}{2 \kappa ^2} \int _{\mathcal{I} ^ + }  du d \theta d \phi \,  \partial _u \left(  \bar{\epsilon} \gamma _5 \gamma _D  \psi _A \, \delta C^D _{B} \, \varepsilon ^{AB} \right)
\end{flalign}
where we have used the fact that $\gamma _5  = \gamma _0 \gamma _1 \gamma _2 \gamma _3 $ and $\gamma _0 \gamma _1 \psi _A ^{(0)} =  - \psi _A^{(0)}$, which is a consequence of $\gamma ^{u(0)} \psi _A ^{(0)} = 0 $. Furthermore, writing this expression as a total derivative is possible because the leading order terms of the spinors and frame field do not depend on time. At subleading order this will no longer be true. 

The overall the charge is,

\begin{flalign}\label{susycharge}
\delta  H _{\epsilon} + i \lambda \cancel{ \delta }\tilde{H} _{\epsilon} =-\frac{1}{ 2\kappa ^2 }  \int _{\partial \Sigma }  \,  \delta \left( \bar{\epsilon } \gamma _5 \gamma \wedge  \psi \right) +\bar{\epsilon } \gamma _5  \delta \gamma \wedge  \  \psi  -   \bar{\epsilon } \gamma _5 \gamma \wedge \delta \psi - i \lambda \ \bar{\epsilon } \, \delta \gamma \wedge \psi  
\end{flalign}

We see that unlike the Rarita-Schwinger case, there is contribution to the magnetic charge. Furthermore because the boundary metric is fixed and because of \ref{conpsi} both charges are finite. We can make both charges integrable by requiring that any non-specific variation (that is not BMS or susy) of $\psi _A ^0$ vanishes to leading order. In the case where  the variation is supersymmetry or super-translation this still holds. When the variation is a super-rotation one may naively think that the charge is not-integrable. However, this is not the case. A simple way to understand why this is false is by noting that $\delta _{\xi} {H} _{\epsilon } = - \delta _{\epsilon} {H} _{\xi}  $. However, we already know that ${H} _{\xi }$ is integrable up to some flux term \cite{27}. Therefore, the obstruction that one gets can only be interpreted as a central charge. This is studied in detail in the last section.

 It is worth noting that the integrable part of the normal charge is the same as the gauge charge of the Rarita-Schwinger field \cite{24} up to an overall numerical constant.  In terms of the metric and gravitino components, the integrable charges are,

\begin{flalign}
 H _{\epsilon} + i \lambda \tilde{H}_{\epsilon}  =\frac{1}{ 2 \kappa ^2 }  \int _{\partial \Sigma }  \,2 \bar{\epsilon } ^{(0)}\gamma _5  \gamma _{D} ^{(1)}\psi _{[A} ^{(0)} \, C^{D} _{\ B]}  -  i \lambda \bar{\epsilon } ^{(0)} \gamma _{D} ^{(1)}\psi _{[A} ^{(0)} \, C^{D} _{\ B]}
\end{flalign}

We see that both charges depend on the tensor $\tilde{C} _{AB} = C^C _{ \ ( B } \varepsilon _{A) C}$, which also enters in the expression for the dual diffeomorphism charges.

\section{Algebra of integrable charges}

Before studying the algebra of the charges, we need to specify the algebra of the generators $\xi$ and $\epsilon$. The algebra for the BMS generators is well-know \cite{17},

\begin{flalign}
&\xi _{f, f^A} = \{ \xi _{f_1. f_1 ^A}, \, \xi_{f_2, f^A_2} \} \\
& f = f^A_1 D_A f_2  - \frac{1}{2} f_1 D_A f^A _2 - (1\leftrightarrow 2) \quad f^A = f^B _1 D_B f^A _2 - (1\leftrightarrow 2)
\end{flalign}

 Furthermore, we already know that the anti-commutator of two supersymmetric parameters  is  a translation, characterized by the function $\bar{\epsilon }_1 \gamma ^{u(0)} \epsilon _2$. The problem now is to work the bracket of $\epsilon $ with $\xi$. We  define the bracket between a BMS vector and the gauge spinor to be,

\begin{flalign}
[\epsilon, \xi _{R}] = \frac{1}{2}\mathcal{L}_{\xi } \epsilon  = f^A \partial _A \epsilon  - \frac{1}{4} D_A f^A \left( \mathbb{1}+\gamma _0 \gamma _1  \right) \epsilon +O(r^{-1})
\end{flalign}

We see that to leading order, this vanishes for a super-translations. Asymptotically the generators of super-translation and supersymmetric transformations commute, as they should.
Using the fact that $\gamma ^{u(0)}  \left( \mathbb{1}+\gamma _0 \gamma _1  \right) = 2 \gamma ^{u(0)} $, one can verify that with this bracket the following Jacobi identity is satisfied.

\begin{flalign}\label{Jacobi}
&\{ \epsilon _1, [\epsilon _2 , \xi _{SR}]  \}- \{ \epsilon _2 , [\epsilon _1 , \xi _{SR}] \}= [\xi _R, \{ \epsilon _1 , \epsilon _2\}] 
\end{flalign}

This result agrees with that in \cite{38}, where the same bracket was established in the super-symmetric extension of BMS.

We will now discuss an interesting result from this bracket. The fact that commutator of gauge spinors can only give an ordinary translation implies that only Lorentz transformations are allowed in the super BMS group. In super gravity we cannot extend the Lorentz group to include arbitrary conformal transformations of the celestial sphere.
In order to understand why this is the case, using the notation of \cite{37}, we write the vector field on the sphere $f^A$, the spherical harmonics $f_{m. n}$ and the covariantly constant spinors $\epsilon $ as,

\begin{flalign}
&  l_n = f^A _n =  -z ^{n+1} \partial _z \quad \bar{l}_n =\bar{f} _n^A = - \bar{z} ^{n+1} \partial _{\bar{z}} \quad f_{m, n} = \frac{1}{ 1+z\bar{z}} z^n \bar{z}^m\\
&\epsilon ^{T} _+ =  e^{i \frac{\phi}{2}} \begin{pmatrix} a e^{ \frac{i\theta}{2}}, & a e^{ - \frac{i\theta}{2} }, & b e^{ \frac{i\theta}{2} }, & b e^{ - \frac{i\theta}{2} } \end{pmatrix} \\
& \epsilon ^{T} _- =  e^{i \frac{-\phi}{2}}\begin{pmatrix} c e^{ \frac{i\theta}{2} }, & -c e^{ - \frac{i\theta}{2} }, & d e^{ \frac{i\theta}{2} }, & -d e^{ - \frac{i\theta}{2} } \end{pmatrix}\\
& f_{0,0} =  \bar{\epsilon }_+ \gamma ^{u(0)} \epsilon _+ + \bar{\epsilon }_- \gamma ^{u(0)} \epsilon _- \quad f_{1,1} =\bar{\epsilon }_+ \gamma ^{u(0)} \epsilon _+ - \bar{\epsilon }_- \gamma ^{u(0)} \epsilon _- \\
&f_{1,0}  = \bar{\epsilon }_- \gamma ^{u(0)} \epsilon _+  \quad f_{0,1} = \bar{\epsilon }_+ \gamma ^{u(0)} \epsilon _-
\end{flalign}

Where $z = e^{i \phi } \cot \frac{\theta }{2} $ is the standard stereographic projection, complex coordinate. the We can re-write the BMS algebra in the following way \cite{37},

\begin{flalign}
&[l_n , l_m] = (n-m) l_{n+m} \quad [\bar{l}_n , \bar{l}_m] = (n-m) \bar{l}_{n+m} \quad [\bar{l}_n , l_m] =0\\
&[l_k, f_{n. m}] = \left( \frac{k+1}{2} -m \right) f_{m+k, n} \quad [\bar{l}_k, f_{n. m}] = \left( \frac{k+1}{2} -n \right) f_{m+k, n}
\end{flalign}

From these equations and the Jacobi identity \eqref{Jacobi} we see that whenever $n > 1$ the commutators of $[l _n, \epsilon]$ and $[\bar{l}_n, \epsilon]$ are equal to spinors, whose components must depend on higher order spin $1/2$ spherical harmonics. This is not allowed by the equations of motion for the gauge spinor. Therefore the allowed asymptotic symmetry group is the super-Poincar\'{e} group, plus the infinite dimensional super-translations.

\subsection{Normal charges}

In this section we study the algebra of the normal Hamiltonian charges, conjugate to diffeomorphic and supersymmetric transformations. They are a representation of an extension of the BMS group - the super-BMS group \cite{26}. It contains a copy of the super-Poincar\'{e} group. The commutators of the BMS group are already well understood \cite{17}. Curiously the presence of the gravitino field does not modify the central charge. Here, we will look only at the commutators, involving supersymmetric charges. 

We first look at the commutator of two supersymmetric charges. This is the same as the commutator for gauge charges of the free Rarita-Schwinger field \cite{31}.

\begin{flalign}
 [ H_{ \epsilon 1}, H_{\epsilon 2} ] =  \frac{1}{\kappa ^2} \int _{\partial \Sigma } \nabla _{[\mu } \bar{\epsilon }_1 \gamma _{\sigma ]} \gamma _5 \epsilon _2 -  (\epsilon _1 \leftrightarrow \epsilon _2) 
\end{flalign}

In \cite{31} it is described in detail how the above expression is a BMS charge, generated by $\xi ^{\mu } =  \bar{\epsilon }_1   \gamma ^{\mu}\epsilon _2  $. This can only be a super-translation, so the commutator is equal to,

\begin{flalign}
[ H_{ \epsilon 1}, H_{\epsilon 2} ] =&\frac{1}{2\kappa ^2}\int _{\partial \Sigma }\sqrt{\gamma } \,  \bar{\epsilon }_1  ^0 \gamma ^{u(0)}\epsilon _2 ^0 \, M -  (\epsilon _1 \leftrightarrow \epsilon _2)  
\end{flalign}

Notice that there is no central charge. What used to be the central charge for this commutator in the free Rarita-Schwinger theory is now the super-translation charge \cite{24}, \cite{31}. As already explained, $\bar{\epsilon }_1  ^0 \gamma ^{u(0)}\epsilon _2 ^0$ is just a lowest order spherical harmonic, and therefore the charge above is a global translation. Again, we note that other supertranslations can not be obtained through the commutator of fermionic charges. 

Now we  proceed to study the commutator of a BMS  and a supersymmetric  charge. Looking at the expressions from subection~\ref{4.3} and section~\ref{2} we can identify the commutator and the central charge as,
 
\begin{flalign}
[H_{\epsilon}, H_{\xi}]= \frac{1}{\kappa ^2} \int _{\partial \Sigma } \, (\mathcal{L}_{\xi}  \bar{\epsilon }) \gamma _5 \gamma \wedge \psi  + \, \bar{\epsilon } ^{(0)}\gamma _5  \gamma _{D} ^{(1)}\psi _{[A} ^{(0)} \, \mathcal{L}_{\xi}C^{D} _{\ B]} = H_{[\epsilon, \xi ]} + K_{ \epsilon , \xi }
\end{flalign}

Notice that we obtain the same result if replace $\delta$, by $\delta _{\xi}$  in \ref{susycharge} and use the act that $\int _{\mathcal{I} ^+ _-}\delta _{\xi } \left(  \bar{\epsilon } \gamma _5 \gamma \wedge \psi \right) = 0$.

For super-translations $\mathcal{L} _{\xi } \epsilon  = O(r^{-1})$, so $H_{[\epsilon, \xi]} = 0$ in agreement with the Poincarr\'{e} algebra. The central charge is,

\begin{flalign}
 K_{\epsilon, \xi _{ST}}  =\frac{i \lambda }{ \kappa ^2}  \int _{\partial \Sigma }  \bar{\epsilon } \gamma _5 \gamma ^{B(-1)} \psi _{[A} ^{(0)} \, D_{D]} D_B f 
\end{flalign}

It is worth nothing that for $l = 0, 1$ the spherical harmonics satisfy $D_A D_B f = \gamma _{AB}f $ and the integral will vanish. Curiously, for higher order spherical harmonics it will not necessarily vanish. This is also due to the fact that, as shown in appendix \ref{C}, $\psi _A$ can depend on arbitrarily high spherical modes. This can be restricted in order to make the central charge vanish.  This expression also have a term involving $\partial _u C_{AB}$. This is the non-integrable part and it vanishes at the past boundary of $\mathcal{I} ^{+}$. The integrable central charge for the super-rotations is,

\begin{flalign}
K_{\epsilon, \xi _{SR}} =   &  \frac{i \lambda}{ \kappa ^2}  \int _{\partial \Sigma }  D_C f^C \, \bar{\epsilon } \gamma _5\gamma ^{B} \psi _{[A}  \, C_{D] B} +   \bar{\epsilon } \gamma _5 \gamma ^{B(-1)} \psi _{[A} ^{(0)}  D_{|C| } f_{D]} C_{B} ^C +  \bar{\epsilon }  \gamma _5 \gamma ^{B(-1)} \psi _{[A} ^{(0)}  C_{D] } ^{C}  D_{[B} f _{C]} 
\end{flalign}

We notice strong a parallel between this central charge and the one for super-rotations, given in equation 7.11 of \cite{16}.

\subsection{Dual charges}

The algebra of the dual charges is simpler to study. The commutator of two dual diffeomorphsim charges is already studied in \cite{16}. It was established that the magnetic BMS charges satisfy the same algebra as the normal ones, but with slightly different central extension. Here we will look at commutators, involving dual fermioinic charges,

\begin{flalign}
 \{ \tilde{H}_{ \epsilon 1}, \tilde{H}_{\epsilon 2} \} = \frac{i \lambda }{2 \kappa ^2} \int _{\partial \Sigma } \nabla _{[\mu } \bar{\epsilon }_1 \gamma _{\sigma ]} \epsilon _2 -  (\epsilon _1 \leftrightarrow \epsilon _2)  = 0 
\end{flalign}
 
This seems to defer from the usual super-Poincar\'{e} algebra. However it is worth nothing, that the dual global translation charge also vanishes \cite{15}. Furthermore, the above expression will not vanish if $C_{AB}$ is not a continuous function on the sphere. This is precisely the condition we need for the existence of BMS magnetic charge as well. We now proceed to compute the commutator of supersymmetric and diffeomorphism charge.

\begin{flalign}
&\{ \tilde{H} _{\epsilon} , \tilde{H}_{\xi } \} =\frac{ i \lambda}{4 \kappa ^2}  \int _{\partial \Sigma } \delta _{\epsilon} \left( e^a \wedge \mathcal{L} _{\xi } e_a \right) - \mathcal{L} _{\xi } \left( e^a \wedge \delta _{\epsilon } e_a \right) = \\
 &  \frac{i \lambda }{2 \kappa ^2}   \int _{\partial \Sigma } \, e^a \wedge \left( \mathcal{L} _{\xi } \epsilon \right) \gamma _a \psi +   \delta _{\epsilon}  e^a \wedge \mathcal{L} _{\xi } e_a   =
  \tilde{H}_{[\epsilon, \xi ]} + \tilde{K}_{\epsilon, \xi } 
\end{flalign}

The first term is the magnetic supersymmetric charge, defined from a gauge spinor $\mathcal{L} _{\xi} \epsilon$. The second one is the central charge. Its expression is almost identical to the one for the normal charges,

\begin{flalign}
& \tilde{K}_{\epsilon, \xi _{ST}}  =\frac{i \lambda }{2 \kappa ^2}  \int _{\partial \Sigma }  \bar{\epsilon } \gamma ^{B(-1)} \psi _{[A} ^{(0)} \, D_{D]} D_B f \\
& \tilde{K}_{\epsilon, \xi _{SR}} =   \frac{i \lambda}{2 \kappa ^2}  \int _{\partial \Sigma }  D_C f^C \, \bar{\epsilon }\gamma ^{B} \psi _{[A}  \, C_{D] B} +   \bar{\epsilon } \gamma ^{B(-1)} \psi _{[A} ^{(0)}  D_{|C| } f_{D]} C_{B} ^C +  \bar{\epsilon } \gamma ^{B(-1)} \psi _{[A} ^{(0)}  C_{D] } ^{C}  D_{[B} f _{C]} 
\end{flalign}

\section{Conclusion}

In this paper, a new type of charges was discovered for $\mathcal{N}=1$ supergravity. They are dual to the usual diffeomorphisms and supersymmetric charges. The diffeomorphism dual charges are the same as the ones studied before \cite{15}, \cite{16}. The supersymmetric magnetic charges are new and do not appear for the free Rarita-Schwinger field \cite{24}. We saw that their integrability and finiteness crucially depend on the conditions on the gravitino field, imposed by the asymptotically flat metric. It would be interesting to study the dual charges for more complicated supergravity theories, or for more exotic metrics. The magnetic charges in supergravity can have potential interesting applications for black holes in supergravity like the extreme Reissner-Nordstr$\ddot{o}$m black hole.

\appendix

\section{Conventions and useful identities}

The choice of  $\gamma $ matrices is,

\begin{flalign}
\gamma ^0  =-i \begin{pmatrix} 0 &\mathbb{1} \\ \mathbb{1} & 0 \end{pmatrix} \quad \gamma ^j  = -i \begin{pmatrix} 0 &  \sigma ^j   \\ - \sigma ^j & 0 \end{pmatrix} \quad  \gamma ^5  =- i \begin{pmatrix} \mathbb{1} &  0  \\ 0 & -\mathbb{1} \end{pmatrix}  \quad C = \begin{pmatrix} 0 &\mathbb{1} \\  - \mathbb{1} & 0 \end{pmatrix} 
\end{flalign}

The covariant and Lie derivative of a spinor are defined as,

\begin{flalign}
\nabla _{\mu } \psi _{\nu } = &\partial _{\mu } \psi _{\mu } - \Gamma ^{\lambda } _{\mu \nu } \psi _{\lambda } + \frac{1}{4} \omega _{\mu a b } \gamma ^{ab } \psi _{\nu } \\
\mathcal{L}_{\xi} \psi _{\nu } =& \xi ^{\mu } \nabla _{\mu } \psi _{\nu } + \psi _{\mu } \nabla _{\nu } \xi ^{\mu } - \frac{1}{4} \nabla _{\mu } \xi _{\rho } \gamma ^{\mu } \gamma ^{\rho } \psi _{\nu }\\
=&  \xi ^{\mu } \partial_{\mu } \psi _{\nu } + \psi _{\mu } \partial _{\nu } \xi ^{\mu } - \frac{1}{4}e_{a [\mu } \mathcal{L} _{\xi }  e^{a} _{\nu ] }\gamma ^{\mu \nu } \psi _{\nu }
\end{flalign}

Up to leading order the last term is simplified by local a Lorentz transformation.

Simple, but useful commutation property,

\begin{flalign}
 \label{5.8}
&  - \lambda _{ab}\gamma ^a \gamma ^b \gamma _5 e_{c \nu } \gamma ^{c}  +  \lambda _{ab} \gamma _5 e_{c\nu } \gamma ^{c}  \gamma ^a \gamma ^b  + 4 \lambda _{ab} e^a _{\nu} \gamma ^b = 0
\end{flalign}

for any $\lambda _{ab}$ that is antisymmetric in $a$ and $b$.

Fierz identity in 4 dimensions \cite{29},

\begin{flalign}\label{fierz}
(\bar{\lambda }_1 \lambda _2 )\lambda _{3 \, a }=& -\frac{1}{4} (\bar{\lambda }_1 \lambda _3 )\lambda _{2 \, a } - \frac{1}{4} (\bar{\lambda }_1 \gamma _5 \lambda _3 ) \gamma _5 \lambda _{2 \ a }  -\frac{1}{4} (\bar{\lambda }_1 \gamma _{\mu} \lambda _3 )(\gamma ^{\mu}\lambda _{2 } )_a \\
&+ \frac{1}{4} (\bar{\lambda }_1 \gamma _{\mu} \gamma _5 \lambda _3 )(\gamma ^{\mu} \gamma _5 \lambda _{2 } )_a + \frac{1}{8} (\bar{\lambda }_1 \gamma _{\mu \nu } \lambda _3 )(\gamma ^{\mu \nu }\lambda _{2 } )_a
\end{flalign}

Cyclic identity \cite{22},

\begin{flalign} \label{cyclic}
\bar{\lambda }_{[1|} \gamma _a \lambda _{|2|}  \bar{\lambda } _{|3]} \gamma ^a
\end{flalign}

Where $\lambda _i$ are arbitrary spinors and the letter $a$ is a spinor index.

\section{Covariantly constant spinor}

The differential equation, satisfied by a covariantly constant spinor $\chi $, is,

\begin{flalign}
\gamma ^A \partial _A \chi + \gamma ^1\chi +  \frac{1}{2} \cot \theta \gamma ^2  \chi = 0
\end{flalign}

With the particular choice we have made for the gamma matrices, the equations for the first and second components are coupled, and are the same as the equations for the third and forth. In particular we have,

\begin{gather}
 i\partial _{\theta }\chi _2 - \frac{1}{\sin \theta } \partial _{\phi } \chi _1 + \frac{i}{2} \cot \theta \chi _2 = \chi _2  \\
 i \partial _{\theta }\chi _1 - \frac{1}{\sin \theta } \partial _{\phi } \chi _2 + \frac{i}{2} \cot \theta \chi _1 =   -\chi _1 
\end{gather}

We assume separation of variables and impose, $\chi _1 =e^{- i \frac{\theta }{2 }} f(\phi )$ and $\chi _2 =e^{ i \frac{\theta }{2 }} h(\phi )$,

\begin{gather}
 e^{- i \frac{\theta }{2 }} \partial _{\phi } f - \frac{i}{2} e^{ - i \frac{\theta }{2 } } h = 0 \quad 
e^{ i \frac{\theta }{2 }} \partial _{\phi } h  -\frac{i}{2} e^{ i \frac{\theta }{2 } } f = 0\\
\Rightarrow \partial _{\phi } h  = \frac{i}{2} f \quad \partial _{\phi } f  = \frac{i}{2} h
\end{gather}

Setting $h=a e^{i \frac{\phi }{2}} + b e^{ - i \frac{\phi }{2}} $ and $f=c e^{i \frac{\phi }{2}} + d e^{ - i \frac{\phi }{2}} $ we have $a=c$ and $b=-d$. 

\section{Equations of motion for $\psi $ }\label{C}

In order to study the the charges and their properties, we need first to solve (to some extent) the equations of motion.Since the metric is already fixed, in this section we need to study only the gravitino field. The equations for motion for it, subject to the gauge condition \ref{fgauge}, are $\gamma ^{\mu } \nabla _{\mu } \psi _{\nu } = 0$.  Taking into consideration \ref{conpsi}, at different orders and for the different components of the gravitino, the exact equations are,

\begin{itemize}

\item{$\psi _A$}:
\begin{flalign}\label{uA}
O(1) : & \gamma ^{u(0)} \partial _u \psi _A =0 \\\label{A2}
O(r^{-1}) : &\gamma ^B \partial _B \psi _A  + \frac{1}{2r} \cot \theta \gamma ^2 \psi _A -  \gamma ^C \Gamma ^B _{CA} \psi _B+ h _{AB} \gamma ^B \psi _{u} +  \gamma ^{u(0)} \partial _u \psi _A  ^{-1} =0
\end{flalign}

\item{$\psi _u$:}
\begin{flalign}
O(r^{-1}) : & \gamma ^{u } \partial _{u } \psi _u  =0 \\ \label{u2}
O(r^{-2}): &  r  \gamma ^A \partial _A \psi _u  + \frac{1}{2} \cot \theta \gamma ^2 \psi _u +   \gamma ^{u } \partial _{u } \psi _u ^{-2} - r \gamma ^A   \partial _u C_{A} ^B \psi _B  =0
\end{flalign}

\item{$\psi _r$:}
\begin{flalign}
 \label{r2}
O(r^{-3}) : & \gamma ^{A(-1)} \partial _A \psi _r ^{(-2)} + \frac{1}{2} \cot \theta \gamma ^2 \psi _r ^{(-2)} - \gamma ^1 \psi _r   - \gamma ^{A(-1)} \psi _A ^{(-1)} \\
& +  \gamma ^{C(-1)} \frac{1}{2} C^{\ A} _{C}  \psi _A ^{(0)} -  \gamma ^{A(-2)}\psi _A ^{(0)} =0 
\end{flalign}

\end{itemize}

We note that $K_{\mu \nu \rho } \gamma ^{\mu } \gamma ^{\nu } \gamma ^{\rho } \sim O(r^{-2})$ and that's why it does not enter these equations.
Using the gauge condition \ref{fgauge} at second order the last equation can be re-written as,

\begin{flalign}\label{C7}
\gamma ^{A(-1)} \partial _A \psi _r ^{(-2)} + \frac{1}{2} \cot \theta \gamma ^2 \psi _r ^{(-2)}  +  \gamma ^{C(-1)} \frac{1}{2} C^{\ A} _{C}  \psi _A ^{(0)}  + \gamma ^{u(0)} \psi _u ^{(-2)}=0 
\end{flalign}

We would need to components of the gravitino field to study the commutators of the charges. This is an integral evaluated at $u \rightarrow - \infty$. We would like the field to be finite in this limit, so we assume it can be decomposed as $\psi _{\mu } = \varphi _{\mu } ( x^A ) + \phi _{\mu } (u, x^A)$ with $\lim _{u \to - \infty } \phi _{\mu } = 0$. Let's look at the equation for $\varphi _u ^{(-1)}$.

\begin{flalign}
\gamma ^A \partial _A \varphi _u  + \frac{1}{2} \cot \theta \gamma ^2 \varphi _u ^{(-1)} = 0
\end{flalign}

Remembering that $\gamma ^{u(0} \varphi _u ^{(-1)}= 0$, this equation is solved by the spinor $\sqrt{\sin \theta } \, \varphi _u ^{(-1)} = \begin{pmatrix}a z ^{ -m}, &a z^{-m}, &- b  \bar{z}^{ - m}, &b \bar{z}^{ - m} \end{pmatrix} ^T$, where $z$ is the stereographic coordinate $z = e^{i \phi } \tan \frac{\theta }{2}$, and $a$ and $b$ are arbitrary constants. Taking into consideration that $\gamma ^{A(-1)} \psi _A ^{(0)}$, the equation for $\varphi _{\phi } ^{(0)}$ is,

\begin{flalign}
\gamma ^A \partial _A \varphi _{\phi}  + \frac{1}{2} \cot \theta \gamma ^2 \varphi _{\phi } ^{(-1)} + \gamma _{\phi } \varphi _u ^{(-1)} = 0
\end{flalign}

If we write $\varphi _{\phi }  = \begin{pmatrix}\iota & \lambda \end{pmatrix}$, the equations for the components of $\lambda$ are,

\begin{flalign}
& \partial _{\theta } (\lambda _1  + \lambda _2)- \frac{i}{\sin \theta } \partial _{\phi } (\lambda _1  + \lambda _2) + \frac{1}{2} \cot \theta  (\lambda _1  + \lambda _2)= 0\\
& \partial _{\theta } (\lambda _1  - \lambda _2)+ \frac{i}{\sin \theta } \partial _{\phi } (\lambda _1  - \lambda _2) + \frac{1}{2} \cot \theta  (\lambda _1  - \lambda _2)= \frac{2b \sin \theta }{\sqrt{\sin \theta }} e^{i m \phi } \cot ^m \frac{\theta }{2}
\end{flalign}

We remember that because $\gamma ^{u(0)} \varphi _A ^{(0)} = 0$, $\lambda _1  + \lambda _2 = 0$. This means that  $ \lambda _1  = -\lambda _2  = \frac{1}{2} \left(\lambda _1 - \lambda _2 \right) = - b \frac{ \cos \theta }{\sqrt{\sin \theta }} z ^{m}$. Similralrly, $\iota _1 = \iota _2 = \frac{1}{2} \left(\iota _1 + \iota _2 \right)   \frac{ \cos \theta }{\sqrt{\sin \theta }} \bar{z} ^{ m}$. From the expression of $\psi _{\phi } $ one can easily work out $\psi _{\theta}$. Curiously they take a form, very similar to the generators of the super-rotations.

\begin{flalign}
&\psi ^{\phi} _m=  \frac{ \cos \theta }{\sqrt{\sin \theta }}  \begin{pmatrix}a z ^{ -m}, &a z^{-m}, &- b  \bar{z}^{ - m}, &b \bar{z}^{ - m} \end{pmatrix} \quad \\
&\psi ^{\theta} _m= i  \sin \theta \frac{ \cos \theta }{\sqrt{\sin \theta }}  \begin{pmatrix}a z ^{ -m}, &a z^{-m}, & b  \bar{z}^{ - m}, &-b \bar{z}^{ - m} \end{pmatrix}
\end{flalign}

Because the algebra of super-rotations has already been established, one can easily work the Lie derivative of this siponor, along a generator of the BMS group (after one corrects with  Local Lorentz transformations,

\begin{flalign}
\mathcal{L}_{l_n + \bar{l}_n } \psi ^A _m = \left(n-m  - \frac{1+\cos ^2 \theta }{2 \cos \theta }\right) \psi ^A _{n+m}
\end{flalign}

It may seem odd that the spinor change under the BMS group, even though the components of the metric, on which its equation of motion depend do not.  However, this linear variation can be compensated by $\delta _{\xi } \psi _u $ and $\delta _{\xi } \psi _A ^{(-1)}$ in the linearized equation of moiton.

From \ref{C7}, we see that the leading order  $u$ - independent part of of $\psi _r$ depends on $\psi _A $ and the tensor $C_{AB}$. For Minkowksi, $\psi _r$ will have the same solution as $\psi _u$. Lastly, $m$ should be a half-integer because $\psi $ is a spinor.
 
\section{ Diffeomorphism variation of the pre-symplectic potential } \label{D}

In this appendix we provide in detail the calculations necessary for the deriving the expression for the diffeomorphism charge.

\begin{flalign}
&2 \kappa ^2 \left(\Theta ( \delta _{\xi })  - \iota _{\xi } L  \right)=\\
 &  \varepsilon _{a b cd} e^a  \wedge  e^b \wedge \mathcal{L} _{\xi} \omega^{ cd } -   \bar{\psi } \wedge  \gamma _5  e^a \gamma _a \wedge \mathcal{L} _{\xi} \psi\\
&-  \iota _{\xi } \left(  \varepsilon _{abcd} e^a \wedge e^b \wedge R ^{cd} - \bar{\psi } \wedge \gamma _5 \gamma \wedge \nabla \psi \right )\\
 & \approx   \varepsilon _{a b c d} e^a \wedge  e^b \wedge \mathcal{L} _{\xi} \omega^{ cd } -  \bar{\psi } \wedge  \gamma _5  e^a \gamma _a \wedge \mathcal{L} _{\xi} \psi \\
&\varepsilon _{a b c d} e^a \wedge  e^b \wedge \mathcal{L} _{\xi} \omega^{ cd } = \varepsilon _{a b c d} e^a \wedge  e^b \wedge\left(  d \iota_{\xi }\omega^{ cd }  + \iota _{\xi } d \omega^{ cd } \right) \\
& \approx d \left( \varepsilon _{a b c d}  e^a \wedge  e^b \iota_{\xi }\omega^{ cd } \right) - 2  \varepsilon _{a b c d}\left( T^a \wedge e^b \iota _{\xi } \omega ^{cd}\right)  + 2  \varepsilon _{a b c d}  \xi ^a   e^b \wedge R^{cd} 
\end{flalign}

 We obtain in the end three terms. One is a total derivative and the other two will be simplified.

\begin{flalign}
&   \bar{\psi } \wedge  \gamma _5  e^a \gamma _a \wedge \mathcal{L} _{\xi} \psi   = \bar{\psi } \gamma _{5} e^a \gamma _{a} \left(  \iota _{\xi } d \psi   +  d \iota _{\xi }  \psi \right)  \\
& \approx  \bar{\psi } \gamma _{5} e^a \gamma _{a}  \iota _{\xi } d \psi   + d \left( \bar{\psi } \gamma _{5} e^a \gamma _{a}  \iota _{\xi }  \psi   \right) - \frac{1}{4} \bar{\psi } \gamma _5 \wedge \omega ^{cd} \gamma _{cd}  \wedge e^a \gamma _{a}  \iota _{\xi }  \psi   +  \bar{\psi } \gamma _{5} e_b \wedge \omega^{ab}  \gamma _{a}  \iota _{\xi } \psi  \\
&- \bar{\psi } \gamma _{5} T^a \gamma _{a}  \iota _{\xi }  \psi   \\
& \approx  \bar{\psi } \gamma _{5} e^a \gamma _{a}  \iota _{\xi } d \psi   + d \left( \bar{\psi } \gamma _{5} e^a \gamma _{a}  \iota _{\xi }  \psi   \right) + \frac{1}{4} \bar{\psi } \gamma _5 \wedge  \gamma \wedge  \gamma _{cd}   \omega ^{cd}  \iota _{\xi }  \psi    - \bar{\psi } \gamma _{5} T^a \gamma _{a}  \iota _{\xi }  \psi   \\
\end{flalign}

Notice that we have omitted the quantity $  \bar{\psi } \wedge  \gamma _5  e^a \gamma _a \wedge  \psi  \, \left(  \omega _{ab} \gamma ^{ab}  -  \nabla _a \xi _b \gamma ^a \gamma ^b \right)$ in the Lie derivative of the gravitino, because it is compensated by the Lorentz transformation.  As we already saw, the charge from this transformation is 0.

\begin{flalign}
 & 2 \varepsilon _{a b c d}  \xi ^a   e^b \wedge R^{cd} =  \bar{\psi } \wedge \gamma _5 \iota _{\xi} \gamma \, \nabla \psi  \approx   \bar{\psi } \wedge \gamma _5 \gamma \, \iota_{\xi}  \nabla \psi   \\
&  2  \varepsilon _{a b c d}\left( T^a \wedge e^b \iota _{\xi } \omega ^{cd}\right)  = \frac{1}{4}\bar{\psi } \wedge \gamma _5 \gamma \gamma_{cd }  \iota _{\xi } \omega ^{cd}  \wedge \psi \\
&   \bar{\psi } \wedge \gamma _5 \gamma \, \iota_{\xi}  \nabla \psi  -  \frac{1}{4}\bar{\psi } \wedge \gamma _5 \gamma \gamma_{cd }  \iota _{\xi } \omega ^{cd}  \wedge \psi  - \frac{1}{4} \bar{\psi } \gamma _5 \wedge  \gamma \wedge  \gamma _{cd}   \omega ^{cd}  \iota _{\xi }  \psi     = \bar{\psi } \gamma _{5} e^a \gamma _{a}  \iota _{\xi } d \psi 
 \end{flalign}

We are left with  $ \bar{\psi } \gamma _{5} T^a \gamma _{a}  \iota _{\xi }  \psi  $  that disappears thanks to the cyclic identity \ref{cyclic}.

\acknowledgments

The author thanks the contribution of her supervisor Malcolm Perry, in motivating, discussing and reviewing
the contents of this paper and Jonathan Crabb\'{e}  and Filipe Miguel for useful feedback. The author is jointly funded by the University of Cambridge, the Cambridge Trust and King’s College.

\end{document}